# Intelligent Tutoring Systems for Generation Z's Addiction

Ioana R. Goldbach
Center for Research and Studies in
Management and Marketing
Valahia University of Targoviste
Targoviste, Dambovita, Romania
e-mail: ioana.goldbach@icstm.ro

Felix G. Hamza-Lup
Computer Science
College of Engineering and Computing
Georgia Southern University
Savannah, Georgia, USA
e-mail: fhamzalup@georgiasouthern.edu

*Abstract*— As generation Z's big data is flooding the Internet through social nets, neural network based data processing is turning an important cornerstone, showing significant potential for fast extraction of data patterns. Online course delivery and associated tutoring are transforming into customizable, on-demand services driven by the learner. Besides automated grading, strong potential exists for the development and deployment of next generation intelligent tutoring software agents. Self-adaptive, online tutoring agents exhibiting "intelligent-like" behavior, being capable "to learn" from the learner, will become the next educational "superstars". Over the past decade, computer-based tutoring agents were deployed in a variety of extended reality environments, from patient rehabilitation to psychological trauma healing. Most of these agents are driven by a set of conditional control statements and a large answers/questions pairs dataset. This article provides a brief introduction on Generation Z's addiction to digital information, highlights important efforts for the development of intelligent dialogue systems, and explains the main components and important design decisions for Intelligent Tutoring System.

## I. INTRODUCTION

Driven by a large amount of data (i.e., training sets) available and the developments in neural-nets, a metamorphosis to *intelligent-like behavior* is catalyzed by the increase in the processing power of parallel systems. Generation Z (or Gen Z, commonly defined as people born between 1995 and mid-2010s) the "digital natives", are becoming more influential in dictating changes in education in the years to come. Like generation Y (i.e., Millennials [1]), Gen Z accelerates the changes in higher education by employing mobile, multimedia, and online technologies. It is the generation of online connection that collaborates and wants to learn fast, adapts and wants active participation in the learning environment. Gen Z students have already entered the university level and they adopt social learning environments that directly involves them. They are the generation of *demanders* as they request services that are available anytime, anywhere. Digital tools are an addiction to many, as they participate on a daily basis in social networking, specifically in scattered cities around the globe where the city architecture is not facilitating social face-to-face interaction.

Research from the Center for Generational Kinetics [2] shows that 95% of the Z generation has smartphones, 55% of them use phones around 5 hours a day, and 26% of them are addicted to digital content, as they spend more than 10 hours a day online. Addiction-like level involvement with digital content shows that 31% of them feel uncomfortable if they are disconnected from the phone for more than 30 minutes. A recent study on people aged between 14 to 40 in the US [3] was targeted towards the behavior, preferences, and attitudes of young people. The study revealed fundamental differences and similarities between the Y and Z generations. About 39% of Gen Z wants to learn with a teacher, while 47% of them spend more than 4 hours a day on video platforms. Compared to Gen Y, Gen Z tends to learn through self-guidance and prefers flexibility. Regardless of the differences between generations, 66% of Gen Z have a positive view of technology in education [3].

This paper is structured as follows: Section 2 provides an overview of an Intelligent Tutoring System (ITS) and the main actors involved in such systems. Section 3 highlights the main research efforts in the area, while the structure and the main components of an ITS are presented in Section 4. In the conclusion, future trends in ITSs evolution are highlighted and explained.

## II. ARTIFICIAL INTELLIGENCE IN EDUCATION SYSTEMS

Artificial Intelligence (AI) mainly resorts to machine learning algorithms to transform data in decisions and provide meaningful user-computer interaction. At the core of the machine learning methodology is a set of statistical and prediction based algorithms or constructs that allow timely big data processing and extraction of meaningful patterns. Such patterns are used to predict (hopefully with high probability, e.g., 90%+) future events/values, hence allowing automated decisions (i.e., expert decision systems) to be taken by machines, providing the user with the impression that the computing device makes intelligent choices.

Particularly interesting is the recent application of AI in intelligent tutoring for education and, as a consequence, the proliferation of ITS. The basic principle of operation and the main actors involved in a possible AI-based ITS are depicted in Figure 1. Data about the learner may be collected from multiple venues (i.e., social networks, instructors, online course preferences, etc.) and recommendations are made based on the processing of collected information and other inputs (e.g., exam results, learner's past and current questions, instructor's feedback about the learner, peers feedback, etc.).

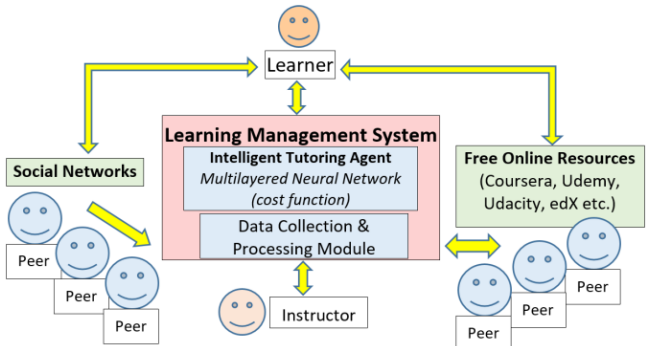

Figure 1. Possible interaction among the actors in future ITS

In the very near future, a data collection and processing module (illustrated in Figure 1) could potentially aggregate information from a variety of sources and could extract patterns specific to the learner, allowing the learner profile generation. Those patterns are further employed by the tutoring system to fine-tune the content of the conversation with the user in order to generate intelligent dialogue. A multilayered neural network, driven by a cost function, is constantly evaluating the learner's feedback and providing informed guidance to the learner.

### III. BRIEF REVIEW ON INTELLIGENT TUTORING

ITSs are not a new development, as early research efforts that focused on intelligent dialogue have been explored for several decades. Among the most notable efforts are the Hamburg Application-oriented Natural language System (HAM-ANS) project [4] at the University of Hamburg, the KLAUS project at the Scientific Research Institute (SRI) International [5] and the XCALIBUR project [6] at Carnegie-Mellon University. Central to these systems was always the requirement for interaction through sequential dialogue with a human operator and the capacity of the system to generate meaningful dialogue based on the collected data.

The rapid proliferation of automated and online learning systems has spawned in the last decade a large number of ITSs with the main goal of enabling the student to successfully solve problems. Among them, AutoTutor [7] is an intelligent guidance system that stimulates dialogue and has the pedagogical strategies of a human tutor. It was designed to help students learn the basics of hardware, operating systems, and the Internet, and enhances the learning technology in the following areas: computer literacy, critical thinking, and physics. AutoTutor focuses on meditation and pedagogical strategies and was designed using human tutor strategies to identify motivational factors for students. AutoTutor was the basis for the development of other intelligent systems such as: AutoManager, AutoTutor-Sensitive, AutoTutor-3D [8] with interactive 3D embedded simulation, DeepTutor, AutoTutor-Lite, GnuTutor, MetaTutor - metacognition self-learning, Human Use Regulatory Affairs Advisor (HURAA) Web Counselor on ethical treatment of experimental subjects, iDRIVE - Learning to Ask Deep Questions about Science, Center for the Study of Adult Literacy (CSAL) and Operation Acquiring Research, Investigative, and Evaluative Skills (ARIES) [9].

Another prominent example is SmartTutor [10] an intelligent system that addresses two basic elements in continuous education: personalization and intelligent guidance. It contains a database of over 3000 reading and math lessons. The effectiveness of the system has been evaluated and the results have been exceptional at the K8 level. The system is based on the fact that learners' answers can provide a lot of information about the current state of their conceptual understanding. The syntactic dimension is explored in Why2-Atlas [11], an intelligent system that analyzes students' explanations of physics principles through various mechanisms. Students introduce their essays into the system as a paragraph, and the tutor uses syntactic analysis to proofread the essays and find misconceptions, as well as incomplete explanations. If the tutor identifies certain mistakes in the essay, it generates a dialogue regarding the wrong or non-existent requirements and then asks the student to correct the essay. Several iterations and dialogues can take place before the process is completed.

Along the same lines, ElectronixTutor [12] is a fully integrated system based on many intelligent learning systems (e.g., AutoTutor, Dragoon, LearnForm, ASSISTments, BEETLE-II). The system includes a student model that has knowledge of electronic circuits and guides other learners in the electronics field providing feedback. Like ElectronixTutor, e-Teacher [13] automatically builds student profiles while studying online courses and detecting the student's performance. The system suggests a customized course of action designed to support each learner.

Introductory knowledge helps learners navigate basic concepts in different disciplines. ZOSMAT [14] has been developed as an intelligent introductory system in response to the student's needs of individual learning. The role of the system is the tracking and the guidance of the learning process. It identifies and records student progress and changes the study program according to the learners' effort. It can be used for individual learning purposes, but it also provides a feature that makes it different from other intelligent guidance systems: it can be used in class under the guidance of a human tutor.

While some of these research efforts are still in the preliminary phases, there are several successful commercial applications, particularly targeted at teaching basic concepts and addressing large groups of learners, specifically at the K-12 level.

## IV. INTELLIGENT TUTORING SYSTEMS STRUCTURE

Intelligent tutoring systems consist of four important components [15]: (1) an Expert Model (EM), (2) a Student Model (SM), (3) a Tutoring Model (TM) and (4) the User Interface Model (UIM), as illustrated in Figure 2. The data flowing among these components is constantly fine-tuned based on the system and the target users group.

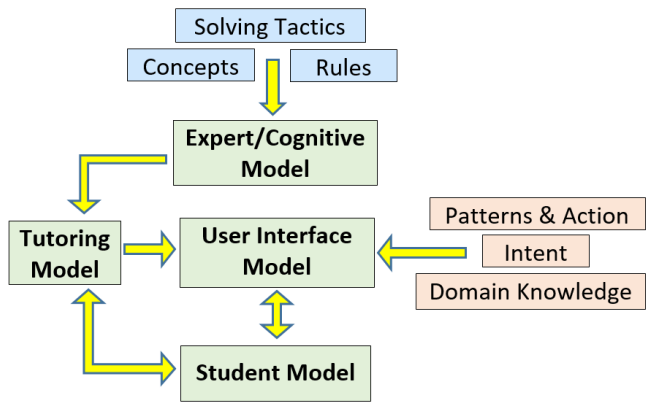

Figure 2. ITS – Generic Structure and Data Flow

The expert model (cognitive/domain model or expert knowledge model) is built on learning theories that consider all the steps required to solve a problem and contains the concepts, rules, and problem-solving tactics of the domain to be learned. The EM also contains the mal-rules and misconceptions that students occasionally exhibit. EM can fulfill several roles: a source of expert knowledge, a standard for evaluating the student's performance or for detecting errors and fallacies. Another approach for developing the EM is the constraint-based modeling approach [16], presented as a set of constraints on correct solutions [17].

The student model can be thought of as a cover on the EM. It is considered as the central component of an ITS, focusing on the student's cognitive and affective states and their evolution as the learning progresses. As the learner works step-by-step through their problem-solving process, the ITS employs a model tracing approach. If the SM deviates from the EM, the system triggers a warning and particular actions. In contrast, in constraint-based tutors, the SM is represented as an overlay on the constraint set [18] and they evaluate the student's solution against the constraint set, to identify satisfied or violated constraints. Violated constraints trigger the ITS feedback on those constraints [19], providing the learner with immediate feedback. The SM builds a profile of strengths and weaknesses for each learner relative to the EM.

Next, the tutor model (or pedagogical model or instructional model) accepts information from the EM and the SM and makes choices about tutoring strategies and actions. The TM contains several hundred production rules that exist in one of two states: learned or unlearned. Every time a student successfully applies a rule to a problem, the system updates a probability estimate that the student has learned the rule. The system continues to drill students on exercises that require the effective application of a rule until the probability that the rule has been learned reaches the 95% threshold [20].

Last but not least, the UIM interprets the learner's contributions through various input media (speech, typing, clicking) and generates output in different media (text, diagrams, animations, agents). It integrates the following information: knowledge about patterns of interpretation (to comprehend the speaker) and action (to generate meaningful expressions) within dialogues, domain knowledge needed for content communication, and knowledge for communicative intent [21]. The communicative intent is the use of gestures, facial expressions, articulations, and/or written expressions to deliver a message and, sometimes, the ITS presents an avatar embodiment to facilitate the user interaction.

ITS are expensive systems to develop both from the complexity and development time perspectives. Attempts to develop authoring tools [22] have looked into various ways to develop agent-based tutors and dialogue-based tutors. Significant research has ensued an array of theoretical frameworks that remain enthusiastically investigated to this day. Reviews of the expert model design in [23]-[25] point to the need to extract domain based features. A review of student modeling [26] reveals the importance of specific learner's characteristics and also points out the requirement for a reward system. A detailed review of tutoring strategies is presented in [27].

Among the most important categorization dimensions for an ITS is the fundamental learning component. Three directions are possible:

- *Simulation-based* learning environments. Here, the general paradigm of a simulated world is captured in the term *reactive environment* [28] to describe an ITS in which the system responds to learners' actions in a variety of ways catalyzing learners' concepts understanding.
- *Discourse-based* learning environments. Natural language interactions have enabled more conversational forms in such environments. Discourse as a tutorial approach, is intended to operate in an ITS much like it does when practiced by a skilled human tutor.
- *Situation-based* learning. Instructional systems may be more effective when coupled with situations in which the users naturally encounter, learn, and apply the skills and knowledge being taught.

A prominent research effort, the Generalized Intelligent Framework for Tutoring (GIFT) [29] is oriented around providing three services: authoring of components, management of instructional processes, and an assessment methodology [30].

## V. Conclusion

The paper presents several statistical facts about Generation Z as it pertains to the use of technology for learning tasks, culminating with the need for customized intelligent tutoring systems. A brief review of the existing ITSs, as well as the fundamental structure of the ITS, is presented with a brief description of each structural component.

The relatively high cost of building an ITS makes it a viable option only for situations such as simultaneous tutoring of large groups, or in cases when tutoring redundancy is necessary and can generate significant savings (i.e., reducing the need for human instructors or freeing human instructors time and resources). With advances in processing speed and machine learning algorithms, we foresee an increase in the online deployment of ITSs and, possibly, a wider adoption of such systems among generation Z's learners.